\documentstyle[aps,twocolumn,eqsecnum]{revtex}
\begin{document}
\draft
\wideabs{
\title{Quantum inequalities in two dimensional Minkowski spacetime}

\author{\'Eanna \'E.\ Flanagan}
\address{Cornell University, Newman Laboratory, Ithaca, NY
14853-5001.}
\maketitle


\begin{abstract}
We generalize some results of Ford and Roman constraining the possible
behaviors of the renormalized expected stress-energy tensor of a free
massless scalar field in two dimensional Minkowski spacetime.  Ford
and Roman showed that the energy density measured by an inertial
observer, when averaged with respect to the observers proper time by
integrating against some weighting function, is
bounded below by a negative lower bound proportional to the reciprocal
of the square of the averaging timescale.  However, the proof required
a particular choice for the weighting function.
We extend the Ford-Roman result in two ways:
(i) We calculate the optimum (maximum possible) lower bound and
characterize the state
which achieves this lower bound; the optimum lower bound differs by a
factor of three from the bound derived by Ford and Roman for their
choice of smearing function.  (ii) We calculate the lower bound for
arbitrary, smooth positive weighting
functions.  We also derive similar lower bounds on the spatial average
of energy density at a fixed moment of time.
\end{abstract}

}

\def\beq{\begin{equation}}
\def\endeq{\end{equation}}
\def\hpl{{{\hat \Phi}_L}}
\def\hpr{{{\hat \Phi}_R}}

\narrowtext

\section{INTRODUCTION AND SUMMARY}
\label{intro}

In classical physics, the energy densities measured by all observers
are non-negative, so that the matter stress-energy tensor $T_{ab}$ obeys
$T_{ab} u^a u^b \ge 0$ for all timelike vectors $u^a$.  This ``weak
energy condition'' strongly constrains the behavior of solutions of
Einstein's field equation:  once gravitational collapse has reached a
certain critical stage, the formation of singularities becomes
inevitable \cite{singul}; traversable wormholes are forbidden \cite{FSW};
and the asymptotic gravitational mass of isolated objects must be
positive \cite{posmass}.

However, as is well known, in quantum field theory the energy density
measured by an
observer at a point in spacetime can be unboundedly negative
\cite{EGJ}.  Examples of situations where observers measure negative
energy densities include the Casimir effect \cite{Casimir}
and squeezed states of light \cite{WKHW}, both of which have been
probed experimentally.  In addition, the theoretical
prediction of black hole evaporation \cite{H75} depends in a crucial
way on negative energy densities.  If nature were to place no
restrictions on negative energies, it might be possible to violate
cosmic censorship \cite{FR90,FR92}, or to produce traversable wormholes
or closed timelike curves \cite{Kip}.  As a consequence,
in recent years there has been considerable interest
in constraints on negative energy density that follow from quantum
field theory.  For reviews of recent results and their ramifications
see, e.g, Refs.~\cite{FR95,FRBH,Yurtsever94,FW}.

In this paper we shall be concerned with so-called ``quantum
inequalities'', which are constraints on the magnitude and duration of
negative energy fluxes and densities measured by inertial observers,
first introduced by Ford \cite{F91} and extensively explored by Ford
and Roman \cite{FR92,FR95,FRBH,FRWH,FRnew}.

\subsection{Quantum Inequalities}

Consider a free, massless scalar field $\Phi$ in two dimensional
Minkowski spacetime.   We consider the following three different
spacetime-averaged observables.
Fix a smooth, strictly positive function $\rho = \rho(\xi)$ with
\beq
\int_{-\infty}^\infty \rho(\xi) d\xi =1,
\label{normalization}
\endeq
which we will call the smearing function.  Let ${\hat T}_{ab}$ be the
stress tensor, and let $(x,t)$ be coordinates such that the metric is
$ds^2 = -dt^2 + dx^2$.  Define
\beq
{\hat {\cal E}}_S[\rho] = \int_{-\infty}^\infty dx \, \rho(x) \, {\hat
T}_{tt}(x,0),
\label{calESdef}
\endeq
\beq
{\hat {\cal E}}_T[\rho] = \int_{-\infty}^\infty dt \, \rho(t) \, {\hat
T}_{tt}(0,t),
\endeq
and
\beq
{\hat {\cal E}}_F[\rho] = \int_{-\infty}^\infty dt \, \rho(t) \,
{\hat T}^{xt}(0,t).
\endeq
The quantity ${\hat {\cal E}}_S[\rho]$ is the spatial average of the
energy density
over the spacelike hypersurface $t=0$, while ${\hat {\cal E}}_T[\rho]$
is the time
average with respect to proper time of the energy density measured by
an inertial observer, and ${\hat {\cal E}}_F[\rho]$ is the time
average with respect
to proper time of the energy flux measured by an inertial observer.
Of these three observables, ${\hat {\cal E}}_S$ and ${\hat {\cal
E}}_T$ are classically positive, while ${\hat {\cal E}}_F$ is
classically positive when only the right-moving sector of the theory
contains excitations.

In the quantum theory, let ${\cal E}_{\rm S,min}[\rho]$ and ${\cal E}_{\rm
T,min}[\rho]$ denote the minimum over all states of the expected value
of the observables ${\hat {\cal E}}_S[\rho]$ and ${\hat {\cal
E}}_T[\rho]$ respectively.
Similarly, let ${\cal E}_{\rm F,min}[\rho]$
denote the minimum over all states in the right moving sector of the
expected value of ${\hat {\cal E}}_F[\rho]$.  Ford and Roman have
previously derived
lower bounds on ${\cal E}_{\rm T,min}[\rho]$ and
${\cal E}_{\rm
F,min}[\rho]$, for a particular choice of the smearing
function $\rho$.  Specifically, they showed that \cite{FR95,FRnew}
\beq
{\cal E}_{\rm T,min}[\rho_0] \ge - {1 \over 8 \pi \tau^2}
\label{r1}
\endeq
and \cite{F91}
\beq
{\cal E}_{\rm F,min}[\rho_0] \ge - {1 \over 16 \pi \tau^2},
\label{r2}
\endeq
where
\beq
\rho_0(t) \equiv {\tau \over \pi} \, {1 \over t^2 + \tau^2}.
\label{rho0def}
\endeq

The main result of this paper is that
\begin{eqnarray}
{\cal E}_{\rm T,min}[\rho] &=& {\cal E}_{\rm S,min}[\rho] = 2 \,
{\cal E}_{\rm F,min}[\rho] \nonumber \\
\mbox{} &=&  - {1 \over 24 \pi}
\int_{-\infty}^\infty dv \, {\rho'(v)^2 \over \rho(v)},
\label{mainresult0}
\end{eqnarray}
for arbitrary smearing functions $\rho(v)$.  Equation
(\ref{mainresult0}) generalizes the Ford-Roman results and shows that
the qualitative nature of those results does not depend on
their specific choice of smearing function (which was chosen to
facilitate the proofs of the inequalities), as one would expect.
Equation (\ref{mainresult0}) also gives
the optimum, maximum possible lower bound on the averaged energy
density, in contrast to the lower bounds (\ref{r1}) and (\ref{r2}).
For the particular choice (\ref{rho0def}) of smearing
function, Eq.~(\ref{mainresult0}) shows that the optimum lower bounds
are a factor of three smaller in absolute value than the bounds
(\ref{r1}) and (\ref{r2}).

Equation (\ref{mainresult0}) holds not just for smearing functions
$\rho(v)$ which are strictly positive [as is the Ford-Roman smearing
function (\ref{rho0def})], but also for smearing functions which are
strictly positive only in an open interval $v_1 < v < v_2$ (with
$v_1$,$v_2$ finite) and zero elsewhere, as long as $\rho(v)$ is smooth on
$-\infty < v < \infty$.
For such smearing functions, the quantity $\rho'(v)^2/\rho(v)$
appearing in Eq.~(\ref{mainresult0}) should be interpreted to be zero
when $\rho(v) = 0$.

Equation (\ref{mainresult0}) also shows that the lower bounds on the
temporal averages and spatial averages of
energy are identical, which is not surprising in a two dimensional
theory.

We derive the result (\ref{mainresult0}) in Sec.~\ref{main} below.
In Sec.~\ref{implications} we discuss some of its implications:
we show that the total amount of negative energy that can be contained
in a finite region $0 \le x \le L$ at a fixed moment of time is
infinite, but that if $\alpha>0$
is a number such that, for some state, $\langle \,{\hat T}_{tt}(x,0) \,
\rangle \le -\alpha$ for all $x$ with $0 \le x \le L$, then $\alpha$
cannot be arbitrarily large.

\section{DERIVATION OF THE QUANTUM INEQUALITY}
\label{main}

We start by showing that the minimum values of the three observables
${\hat {\cal E}}_S$, ${\hat {\cal E}}_T$, and ${\hat {\cal E}}_F$
that we have defined are not independent of each other, {\it c.f.},
the first part of Eq.~(\ref{mainresult0}) above.  To see this, introduce null
coordinates $u = t + x$, $v = t - x$, so that the field operator can be
decomposed as
\beq
{\hat \Phi}(x,t) = \hpr(v) + \hpl(u).
\endeq
Here $\hpr(v)$ acts on the right-moving sector and $\hpl(u)$ on the
left-moving sector of the theory.  The non-zero components of the
stress tensor in the $(u,v)$ coordinates are
${\hat T}_{uu}(u) = :
(\partial_u \hpl)^2 :$ and
\beq
{\hat T}_{vv}(v) = :(\partial_v \hpr)^2 :,
\label{normalorder}
\endeq
where the colons denote normal ordering.
Define the right-moving and left-moving energy flux observables
\beq
{\hat {\cal E}}^{(R)}[\rho] \equiv \int dv \, \rho(v) \, {\hat T}_{vv}(v)
\label{calER}
\endeq
and
\beq
{\hat {\cal E}}^{(L)}[\rho] \equiv \int du \, \rho(u) \, {\hat
T}_{uu}(u).
\endeq
Then we have ${\hat {\cal E}}_S[\rho] = {\hat {\cal E}}_T[\rho] =
{\hat {\cal E}}^{(R)}[\rho] + {\hat {\cal E}}^{(L)}[\rho]$, while
${\hat {\cal E}}_F[\rho] =
{\hat {\cal E}}^{(R)}[\rho] - {\hat {\cal E}}^{(L)}[\rho]$.  It
follows that ${\cal E}_{\rm T,min}[\rho]
= {\cal E}_{\rm S,min}[\rho] = 2 {\cal E}_{\rm F,min}[\rho]  =2 {\cal
E}^{(R)}_{\rm min}[\rho]$, from
which the first part of Eq.~(\ref{mainresult0}) follows.

Thus, to establish Eq.~(\ref{mainresult0}) it is sufficient to
consider the right-moving sector of the theory and to show that
\beq
{\cal E}^{(R)}_{\rm min}[\rho]  = - {1 \over 48 \pi}
\int_{-\infty}^\infty dv \, {\rho'(v)^2 \over \rho(v)},
\label{mainresult}
\endeq
where
\beq
{\cal E}^{(R)}_{\rm min}[\rho] \equiv \min_{\rm states}
\,\,\,\langle {\hat {\cal E}}^{(R)}[\rho] \rangle.
\label{calERdef}
\endeq
We derive the result (\ref{mainresult}) in this section in two stages.
First, in subsection \ref{bog}, we give a simple derivation which is valid
only for smearing functions which are strictly positive and for which the
minimum over states in Eq.~(\ref{calERdef}) is achieved by a state in
the usual Hilbert space [{\it c.f.}, Eq.~(\ref{condt}) below].  Then, in
subsection \ref{alg}, we use the algebraic formulation of quantum
field theory to extend the proof to more general smearing
functions.

\subsection{Bogolubov transformation}
\label{bog}

The key idea in our proof is to make a Bogolubov transformation which
transforms the quadratic form (\ref{calER}) into a simple form.  In
general spacetimes such a Bogolubov transformation is difficult to
obtain, but in flat, two dimensional spacetimes it can be obtained
very simply by using a coordinate transformation, as we now explain.

We can write the mode expansion of the right-moving field operator as
\beq
\hpr(v)  = \, {1 \over \sqrt{2 \pi}}  \int_0^\infty d \omega \, {1 \over
\sqrt{2 \omega}} \, \left[ e^{-i \omega v } {\hat a}_\omega + {\rm
h.c.} \right],
\endeq
where ${\rm h.c.}$ means Hermitian conjugate.  The Hamiltonian of the
right-moving sector is
\beq
{\hat H}_R = \int_0^\infty d \omega \, \omega \, {\hat a}_\omega^\dagger
{\hat a}_\omega.
\label{Hamiltonian}
\endeq
Consider now a new
coordinate $V$ which is a monotonic increasing
function of $v$,
\beq
V = f(v)
\label{Vfv}
\endeq
say, where the function $f$ is a bijection from the real line to itself.
We define a mode expansion with respect to the $V$ coordinate \cite{note0}:
\begin{eqnarray}
\FL
\hpr(v) &=& \hpr[f^{-1}(V)] \nonumber \\
\mbox{} &=& {1 \over \sqrt{2 \pi}} \int_0^\infty d \omega \, {1 \over
\sqrt{2 \omega}} \, \left[ e^{-i \omega V } {\hat b}_\omega + {\rm
h.c.} \right].
\end{eqnarray}
Since the function $f$ is a bijection, the algebra spanned by the
operators ${\hat a}_\omega$ coincides with
the algebra spanned by the operators ${\hat b}_\omega$.  [In
subsection \ref{alg} below we will consider the case where $f$ is a
bijection from a finite open interval $(v_1,v_2)$ to the real line,
and where correspondingly the operators ${\hat a}_\omega$ and ${\hat
b}_\omega$ span different algebras.]  Thus,
the operators ${\hat b}_\omega$ can be expressed as linear
combinations of the ${\hat a}_\omega$'s and ${\hat
a}_\omega^\dagger$'s, and conversely.

We now assume that there exists a unitary operator ${\hat S}$ such that
\beq
{\hat S} \, {\hat a}_\omega  {\hat S}^\dagger = {\hat b}_\omega.
\label{Sdef}
\endeq
Such an operator will not always exist, as we discuss in
Sec.~\ref{alg} below, but for the remainder of this subsection we will
restrict attention to smearing functions $\rho(v)$ for which the
operator ${\hat S}$ does exist.
It follows from Eq.~(\ref{Sdef}) that
\beq
{\hat S}^\dagger \, \hpr(v) \, {\hat S} = \hpr[f(v)].
\label{tf}
\endeq
Consider now the transform
${\hat S}^\dagger {\hat T}_{vv}(v) {\hat S}$ of the operator ${\hat
T}_{vv}(v)$.  Using Eq.~(\ref{normalorder}) this can be written as
\beq
{\hat S}^\dagger {\hat T}_{vv}(v) {\hat S} = \lim_{{\bar v} \to v} \,\,
{\hat S}^\dagger \partial_{\bar v} \partial_v
\left[ \hpr({\bar v}) \hpr(v) - H(v - {\bar v})
\right] {\hat S},
\label{e1}
\endeq
where
\beq
H(\Delta v) = - {1 \over 4 \pi} \left[ \ln|\Delta v| + \pi i
\Theta(-\Delta v) \right]
\label{Hdef}
\endeq
is the distribution that the normal ordering procedure effectively
subtracts off.  Here $\Theta$ is the step function.  Equations
(\ref{normalorder}), (\ref{Vfv}), (\ref{tf}) and (\ref{e1}) now yield
\begin{eqnarray}
{\hat S}^\dagger {\hat T}_{vv}(v) {\hat S} &=& \lim_{{\bar v} \to v} \,\,
\partial_{\bar v} \partial_v \left[ \hpr[f({\bar v})] \hpr[f(v)] - H(v
- {\bar v})
\right] \nonumber \\
\mbox{} &=& \lim_{{\bar v} \to v} \,\,
V'(v)^2 \partial_{{\bar V}} \hpr({\bar V}) \partial_V \hpr(V) \nonumber \\
\mbox{} && \ \
- \partial_{{\bar v}} \partial_v H(v - {\bar v})  \nonumber \\
\mbox{} &=& V'(v)^2 :[\partial_V \hpr(V)]^2 : \,\,- \Delta(v),
\nonumber \\
\mbox{} &=& V'(v)^2 {\hat T}_{vv}(V) \,\,- \Delta(v),
\label{e2}
\end{eqnarray}
where primes denote derivatives with respect to $v$ and
\beq
\FL
\Delta(v) = \lim_{{\bar v} \to v} \, \partial_v \partial_{{\bar v}}
\bigg\{ H(v - {\bar v}) - H[f(v) - f({\bar v})] \bigg\}.
\endeq
Using Eq.~(\ref{Hdef}) we find
\begin{eqnarray}
\Delta(v) &=& {1 \over 4 \pi} \left[ {V'''(v) \over 6 V'(v)} - {V''(v)^2
\over 4 V'(v)^2} \right] \nonumber \\
\mbox{} &=& - {1 \over 12 \pi} \sqrt{V'(v)} \left( {1 \over
\sqrt{V'(v)}} \right)^{''}.
\label{Deltadef}
\end{eqnarray}

The relation (\ref{e2}) is the key result that we shall use.  Note that
taking the expected value of Eq.~(\ref{e2}) in the vacuum state yields
\beq
\left< \psi \right| {\hat T}_{vv}(v) \left| \psi \right> = - \Delta(v),
\endeq
where $\left| \psi \right> = {\hat S} \left| 0 \right>$ is the natural
vacuum state associated with the $V$ coordinate, which satisfies ${\hat
b}_\omega \left| \psi \right> =0$.  This reproduces the standard
formula for the expected stress tensor in the vacuum state associated
with a given null coordinate, see, e.g., Ref.~\cite{standard}.

Now integrate Eq.~(\ref{e2}) against the smearing function $\rho(v)$.
{}From Eq.~(\ref{calER}) this yields
\begin{eqnarray}
{\hat S}^\dagger {\hat {\cal E}}^{(R)}[\rho] {\hat S} &=& \int dv
\rho(v) \, V'(v)^2 {\hat T}_{vv}[V(v)] \nonumber \\
\mbox{} &&- \int dv \rho(v) \Delta(v).
\label{almost}
\end{eqnarray}
We now choose the coordinate $V$ to be such that $\rho(v) V'(v)
=1$; note that this prescription yields a bijection $v \to V(v)$ since $\rho(v)
>0$.  The first term on the right hand side of Eq.~(\ref{almost})
now becomes $\int dV {\hat T}_{vv}(V)$, which is just the Hamiltonian
${\hat H}_R$, c.f. Eq.~(\ref{Hamiltonian}) above.  Inserting the relation
$V'(v) = 1/\rho(v)$ into Eqs.~(\ref{Deltadef}) and (\ref{almost})
gives
\beq
{\hat S}^\dagger {\hat {\cal E}}^{(R)}[\rho] {\hat S} = {\hat H}_R -
\Delta,
\label{finalans}
\endeq
where
\begin{eqnarray}
\Delta  &=& - {1 \over 12 \pi} \int dv \sqrt{\rho(v)}  \left(
\sqrt{\rho(v)}\right)^{''} \nonumber \\
\mbox{} &=& {1 \over 48 \pi} \int dv \, {\rho'(v)^2 \over \rho(v)}.
\label{finalans1}
\end{eqnarray}
On the second line we have integrated by parts, and have assumed that
$\rho'(v) \to 0$ as $v \to \pm \infty$.

It is clear from Eq.~(\ref{finalans}) that ${\cal E}_{\rm
min}^{(R)}[\rho] = -\Delta$, since ${\hat H}_R$ is a positive operator
with minimum eigenvalue zero.  Equation (\ref{mainresult}) then follows
from Eq.~(\ref{finalans1}).  Also, the state which achieves the
minimum value $- \Delta$ of ${\hat {\cal E}}^{(R)}[\rho]$ is just the vacuum
state $\left| \psi \right> = {\hat S} \left| 0 \right>$ associated
with the $V$ coordinate; this is a generalized (multi-mode) squeezed
state.  The $V$ coordinate is given in terms of $\rho(v)$ by
\beq
V(v) = \int {dv \over \rho(v)}.
\label{newcoord}
\endeq

\subsection{Algebraic reformulation}
\label{alg}

The derivation just described suffers from the limitation that
in certain cases the ``scattering matrix'' ${\hat S}$ will fail to
exist.  This operator ${\hat S}$ will exist when \cite{Waldbook}
\beq
\int_0^\infty d\omega \, \int_0^\infty d\omega^\prime \, | \beta_{\omega
\omega'}|^2 < \infty,
\label{condt}
\endeq
where
\beq
\beta_{\omega \omega'} =  \int_{-\infty}^\infty dv {\left[ \omega -
\omega' V'(v)\right] \over \sqrt{ \omega \omega'}} \, e^{-i \omega v}
\, e^{-i \omega' V(v)}.
\endeq
The condition (\ref{condt}) will be violated unless $|V'(v) - 1| < 1$
everywhere, i.e., unless
\beq
\rho(v) > 1/2
\label{condt1}
\endeq
everywhere.  Therefore, for
smearing functions which satisfy the normalization condition
(\ref{normalization}), the Bogolubov transformation to the mode basis
associated with the new coordinate (\ref{newcoord}) does not yield a
well defined scattering operator ${\hat S}$.  The
proof outlined in Sec.~\ref{bog} above is valid only for
non-normalizable smearing functions satisfying (\ref{condt1}).

However, it is straightforward to generalize the proof to
smearing functions for which the condition (\ref{condt}) is violated using
the algebraic formulation of quantum field theory \cite{Waldbook}, as
we now outline.   The following proof also applies to smearing
functions which are strictly positive in an open region $v_1 < v < v_2$
(with $v_1$ and/or $v_2$ finite) and which vanish outside that open
region.  For any algebraic state
$\eta$ on Minkowski spacetime, let
\beq
{\cal F}_{g,\eta}(v) = \langle T_{vv}(v) \rangle_{\eta}
\endeq
denote the expected value of the $vv$ component of the stress tensor
in the state $\eta$.
Here $g = g_{ab}$ denotes the flat Minkowski metric
\beq
g_{ab} dx^a dx^b= -dt^2 + dx^2 = - du dv,
\endeq
where $x^a = (x,t)$.  Now suppose that $V$ is a coordinate on the open
interval $(v_1,v_2)$ [which may be $(-\infty, \infty)$] given by $V =
f(v)$, where $f$ is a monotonically increasing bijection from
$(v_1,v_2)$ to $(-\infty,\infty)$.
Consider the metric ${\bar g}_{ab}$ which is conformally related to
$g_{ab}$ given by
\beq
{\bar g}_{ab} dx^a dx^b = -du dV = - V'(v) du dv.
\endeq
This metric is defined on the submanifold ${\bar M}$ of the original
spacetime defined by the inequality $v_1 < v < v_2$; the pair $({\bar
M}, {\bar g}_{ab})$ is itself a two dimensional Minkowski spacetime.

We can naturally associate with the state $\eta$ on Minkowski
spacetime $(M,g_{ab})$ a state ${\bar \eta}$ on the spacetime $({\bar
M},{\bar g}_{ab})$ which has the same $n$ point distributions $\langle
\hpr(v_1) \ldots \hpr(v_n) \rangle$.  It can be checked that the resulting
algebraic state ${\bar \eta}$ obeys the Hadamard and positivity
conditions on the spacetime $({\bar M},{\bar g}_{ab})$ and so is a
well defined state.  If we define
\beq
{\cal F}_{{\bar g},{\bar \eta}}(v) = \langle T_{vv}(v) \rangle_{{\bar
\eta}},
\endeq
then a straightforward point-splitting computation exactly analogous
to that outlined in Sec.~\ref{bog} above yields
\beq
{\cal F}_{g,\eta}(v) = V'(v)^2 {\cal F}_{{\bar g},{\bar \eta}}[V(v)] -
\Delta(v),
\endeq
where $\Delta(v)$ is the quantity defined by Eq.~(\ref{Deltadef})
above.  Now choosing $V'(v) = 1/\rho(v)$ yields, in an obvious
notation,
\beq
\langle {\hat {\cal E}}^{(R)}[\rho] \rangle_\eta = \langle {\hat H}_R
\rangle_{\bar \eta} - \Delta,
\endeq
where $\Delta$ is given by Eq.~(\ref{finalans1}) but with the domain
of integration being $(v_1,v_2)$.
Finally we use the fact that the quadratic form ${\hat H}_R$ is
positive indefinite for all algebraic states ${\bar \eta}$ (not just for
states in the folium of the vacuum state).  The remainder of the proof
now follows just as before.

\section{IMPLICATIONS}
\label{implications}

In this section we discuss some of the implications of our result
(\ref{mainresult0}).  First, it is possible to deduce from
Eq.~(\ref{mainresult0}) constraints on
the {\it maximum} energy density rather than the averaged energy
density in a region of space.  Specifically, the
quantity
\beq
\min_{\rm states}
\,\,\, \max_{0 \le x \le L} \,\, \langle {\hat T}_{tt}(x,0)
\rangle
\label{obs1}
\endeq
is bounded below for any $L > 0$, which confirms in this context a
conjecture made in Ref.~\cite{FW}.  To see that the quantity (\ref{obs1})
is bounded below, note that
${\hat T}_{tt}(x,t=0) = {\hat T}_{uu}(u=x) + {\hat T}_{vv}(v = -x)$,
so that
\begin{eqnarray}
\max_{0 \le x \le L} \,\, \langle {\hat T}_{tt}(x,0) \rangle
&\le&  \max_{0 \le u \le L} \,\, \langle {\hat T}_{uu}(u) \rangle
\nonumber \\
&&+ \max_{-L \le v \le 0} \,\, \langle {\hat T}_{vv}(v) \rangle.
\label{lrsplit}
\end{eqnarray}
Thus, it is sufficient to bound each term on the right hand side of
Eq.~(\ref{lrsplit}).  Next, for any state, and for any smearing
function $\rho(v)$ with support in $[-L,0]$ and normalized according
to Eq.~(\ref{normalization}), we have
\beq
\langle \, {\hat {\cal E}}_R[\rho] \, \rangle \le \max_{-L \le v \le
0} \, \langle T_{vv}(v) \rangle.
\label{basicineq}
\endeq
One can write down a similar inequality for the other term on the right
hand side of Eq.~(\ref{lrsplit}).  Taking the minimum over states and
using Eqs.~(\ref{mainresult}), 
(\ref{lrsplit}) and (\ref{basicineq}) now yields
\begin{equation}
\min_{\rm states}
\,\,\, \max_{0 \le x \le L} \,\, \langle {\hat T}_{tt}(x,0)
\rangle \ge 2 \max_\rho \, {\cal E}_{\rm R,min}[\rho],
\label{result2}
\end{equation}
where the maximum is taken over all smooth normalizable smearing
functions $\rho$ with support in $[0,L]$.  
It is clear on dimensional
grounds that the right hand side of Eq.~(\ref{result2}) is
proportional to $- \hbar / L^2$, and hence we obtain
\begin{equation}
\min_{\rm states}
\,\,\, \max_{0 \le x \le L} \,\, \langle {\hat T}_{tt}(x,0)
\rangle \ge -k {\hbar \over L^2},
\label{result2a}
\end{equation}
for some constant $k$.

The second implication of our result is that
the {\it total} amount of negative energy that can be
contained in a finite region $0 \le x \le L$ of space in two
dimensions is infinite.  This can be seen from our result applied to
the observable ${\hat {\cal E}}_S[\rho]$,
by taking the limit where the smearing function $\rho(x)$ approaches
the function
\beq
\rho_{\rm box}(x) = \left\{ \begin{array}{ll} 1 &
			\mbox{\ \ $ 0 \le x \le L$,}\nonumber \\
			0 & \mbox{\ \ otherwise}. \\ \end{array} \right.
\endeq
In this limit the quantity ${\cal E}_{\rm S,min}[\rho]$ diverges.
However, this divergence is merely an ultraviolet
edge-effect, in the sense that states which have large total negative
energies inside the finite region will have most of the energy density
concentrated near the edges at $x=0$ and $x=L$ [this can be seen from
Eq.~(\ref{result2a})], and furthermore such states will have
compensating large positive energy densities just outside the finite
region.

\section{CONCLUSION}

We have derived a very general constraint on the behavior of
renormalized expected stress tensors in free field theory in two
dimensions, generalizing earlier results of Ford and Roman
\cite{extension}.  Our
result confirms the generality of the Ford-Roman time-energy
uncertainty-principle-type relation \cite{FR95}: that the amount
$\Delta E$ of
energy measured over a time $\Delta t$ is constrained by
\beq
\Delta E \agt - {\hbar \over \Delta t}.
\endeq
We also showed that the total energy in a one dimensional box is
unbounded below, but that the maximum energy density in such a box is
bounded below.

\acknowledgments
The author thanks Robert Wald, Tom Roman and Larry Ford for helpful
discussions.   This research was supported in part by NSF grants
PHY 9514726 and PHY 9408378, and by an Enrico Fermi fellowship.


\begin{references}
\bibitem{note0} Note that we are using a notation where the argument
of $\hpr$ is always the $v$-coordinate of the spacetime point, never the $V$
coordinate; in other words we shall {\it not} use the notation
$\hpr(V)$ to denote the field operator at the spacetime point with
$V$-coordinate value $V$ and $v$ coordinate value $f^{-1}(V)$.
\bibitem{extension}  Our result can also be applied to deduce lower
bounds on very general space-time smearings of the stress energy
tensor using the fact that for any smooth tensor $f^{ab}(x,t)$, we
have
\begin{eqnarray}
\int dx \, dt \, f^{ab}(x,t) {\hat T}_{ab}(x,t) &=& \int du \, {\hat
T}_{vv}(v) f(v) \nonumber \\
&&+ \int dv \, {\hat T}_{uu}(u) g(u), \nonumber
\end{eqnarray}
where $f(v) \equiv \int f^{vv}(u,v) du$ and $g(u) \equiv \int f^{uu}(u,v)
dv$.  The expected value of two terms on the right hand side can be
bounded below using Eq.~(\ref{mainresult}).
\end{references}
\end{document}